\documentclass[sigconf, 10pt, nonacm]{acmart}
\usepackage[utf8]{inputenc} %
\usepackage[T1]{fontenc}    %

\usepackage{microtype}
\usepackage{graphicx}
\usepackage{xspace}
\usepackage{booktabs} %
\usepackage{subcaption}
\usepackage{enumitem}%
\usepackage{siunitx}

\usepackage{gensymb}

\usepackage{amsmath}
\usepackage{mathtools}
\usepackage{amsthm}
\usepackage{soul}
\usepackage{xcolor}
\usepackage{nicefrac}       %
\usepackage{microtype}      %
\usepackage{makecell}
\usepackage{float}
\usepackage{dsfont} %

\usepackage[capitalize,noabbrev]{cleveref}
\crefformat{section}{#2\S#1#3}
\crefname{equation}{Eq.}{Eqs.}
\Crefname{equation}{Eq.}{Eqs.}
\crefname{figure}{Fig.}{Figs.}
\Crefname{figure}{Fig.}{Figs.}
\crefname{table}{Tab.}{Tabs.}
\Crefname{table}{Tab.}{Tabs.}

\newcommand{\eg}{{\it e.g.,}\xspace}
\newcommand{\ie}{{\it i.e.,}\xspace}

\newcommand{\myspace}{\vspace{1.5pt}} %

\usepackage[style=base,textfont={small,it},belowskip=0pt]{caption}

\newcommand{\myparagraph}[1]{\noindent{\bfseries #1}}

\usepackage{ifthen}
\usepackage{comment}
\newif\ifcomments
\commentsfalse
\ifcomments
\usepackage[textsize=small,color=lightgray]{todonotes}
\paperwidth=\dimexpr \paperwidth + 5cm\relax
\oddsidemargin=\dimexpr\oddsidemargin + 2.5cm\relax
\evensidemargin=\dimexpr\evensidemargin + 2.5cm\relax
\marginparwidth=\dimexpr \marginparwidth + 2.5cm\relax
\else
\usepackage[disable]{todonotes}
\fi

\newcommand{\peter}[1]{\todo[color=cyan!40]{\textbf{PS}: #1}\ignorespaces}
\newcommand{\alex}[1]{\todo[color=magenta!40]{\textbf{AK}: #1}\ignorespaces}

\def\titletext{Managing Bandwidth: The Key to Cloud-Assisted Autonomous Driving}
\title{\fontsize{14pt}{16pt}\selectfont \titletext}

\def\affilbear{\textsuperscript{\includegraphics[height=0.7em]{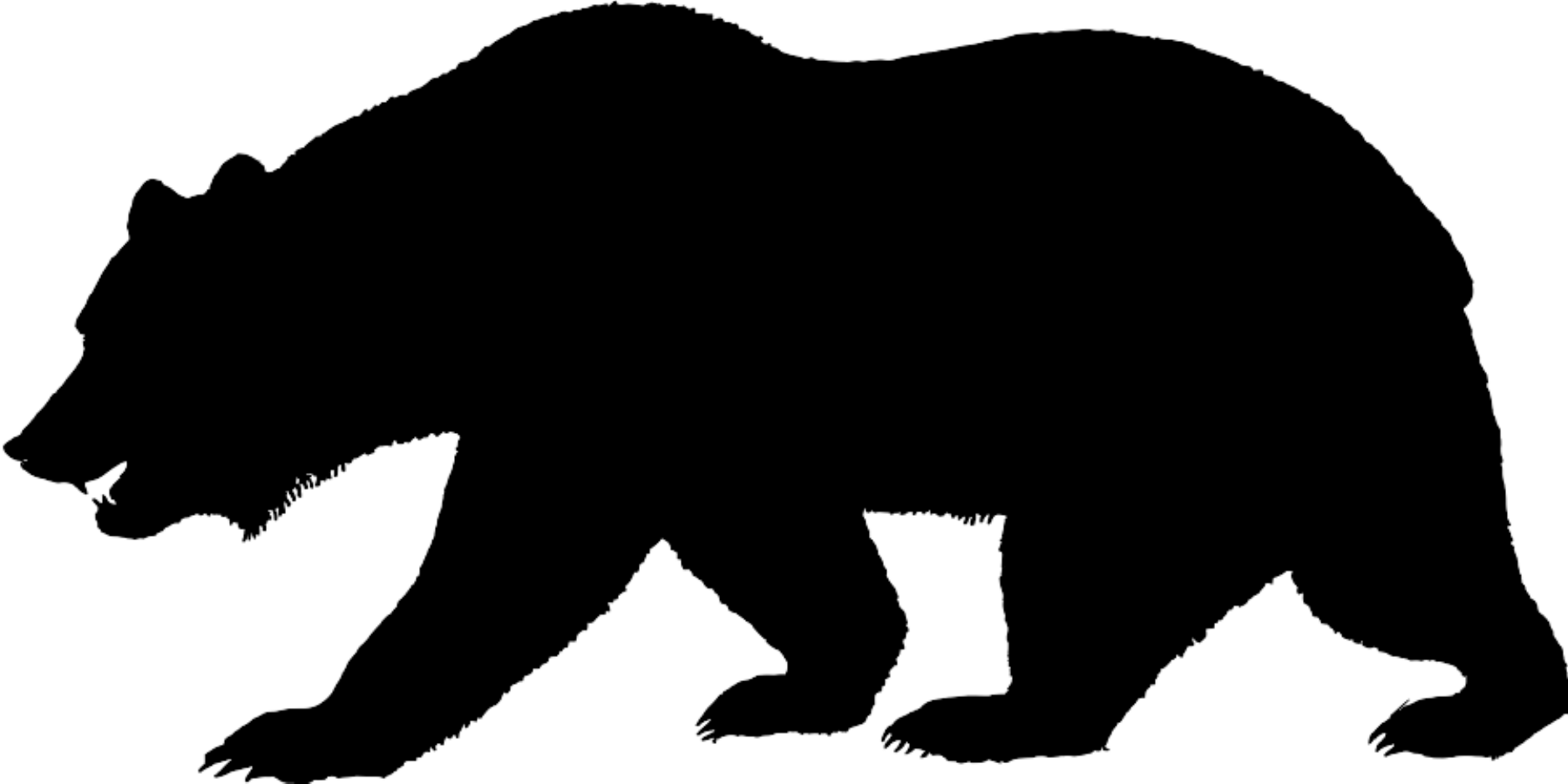}}}
\def\affilicsi{\textsuperscript{\includegraphics[height=0.7em]{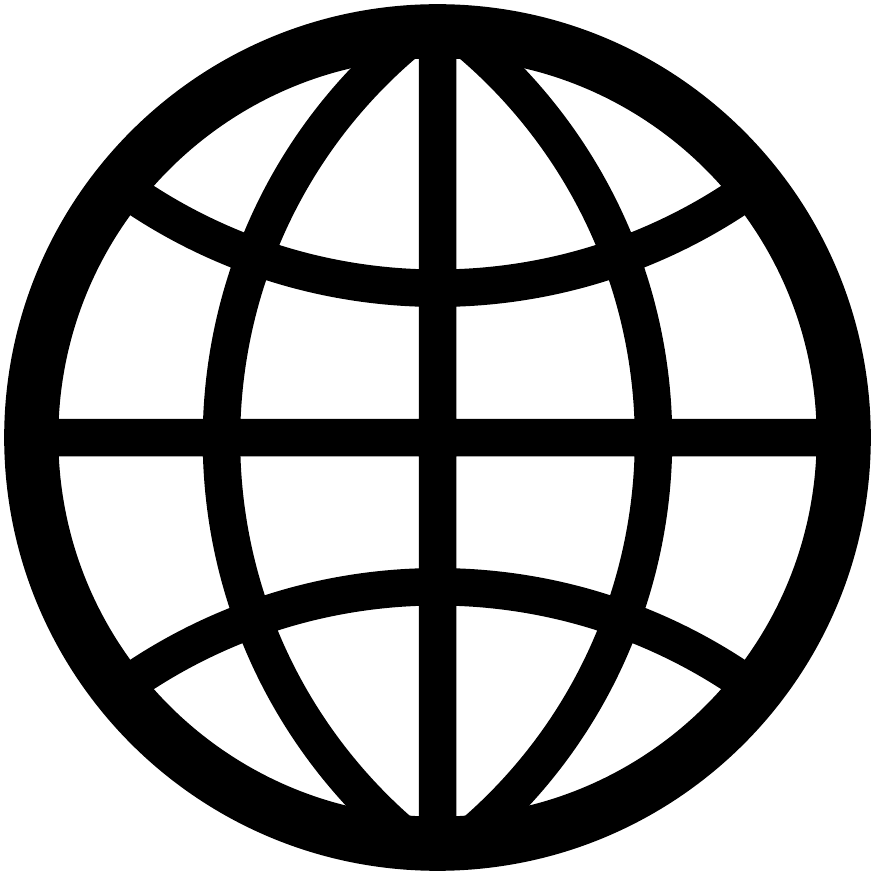}}}
\def\affilscott{\textsuperscript{\includegraphics[height=0.7em]{icons/bear}\includegraphics[height=0.7em]{icons/globe}}}

\author{
  \large
  \begin{tabular}{ccc}
    Alexander Krentsel\textsuperscript{1}\affilbear &
    Peter Schafhalter\textsuperscript{1}\affilbear &
    Joseph E. Gonzalez\affilbear \\
    Sylvia Ratnasamy\affilbear & Scott Shenker\affilscott
    & Ion Stoica\affilbear
  \end{tabular}
}

\affiliation{
  \institution{
    \begin{tabular}{cc}
      \affilbear UC Berkeley & \affilicsi ICSI
    \end{tabular}
  }
  \city{}
  \country{}
}
\renewcommand{\shortauthors}{Alexander Krentsel, Peter Schafhalter, Joseph
E. Gonzalez, Sylvia Ratnasamy, Scott Shenker, and Ion Stoica}
\renewcommand{\shorttitle}{\titletext}

\begin{document}

\begin{abstract}
  Prevailing wisdom asserts that one cannot rely on the cloud for critical
  real-time control systems like self-driving cars.
  We argue that we can, and must.
  Following the trends of increasing model sizes, improvements in hardware,
  and evolving mobile networks, we identify an opportunity to offload
  parts of time-sensitive and latency-critical compute to the cloud.
  Doing so requires carefully allocating bandwidth to meet strict latency
  SLOs, while maximizing benefit to the car.
\end{abstract}

\maketitle
\refstepcounter{footnote}
\footnotetext{Both authors contributed equally to this work.}

\section{Introduction}
\label{s:introduction}

Autonomous driving has the potential to transform society by reducing
road fatalities through the elimination of human
error~\cite{nhtsa-sae-automation},
freeing up to one billion hours spent in traffic per day by improving
traffic flow~\cite{mckinsey-50mins}, 
and providing mobility to millions of people impacted by
disabilities~\cite{claypool2017self}.
While limited deployments of autonomous vehicles (AVs) are
underway~\cite{waymo-scaling-to-four-cities}, challenges remain such as
operation in poor weather conditions and construction
zones~\cite{bloomberg-self-driving-is-going-nowhere}.

To address these challenges, significant efforts have focused on improving the
accuracy of the machine learning (ML) models underpinning autonomous
driving~\cite{chauffeurnet,wu2023point,shi2024mtr++,leng2024pvtransformer,mu2024most}.
However, more accurate models are typically more
compute intensive~\cite{sevilla2022compute,zhai2022scaling}.
Because AVs must operate with faster-than-human reaction times (\eg $390$
ms to $1.2$ s~\cite{wolfe2020rapid,johansson1971drivers}), deploying models
on-vehicle requires a careful navigation of the tradeoffs between runtime
and accuracy to ensure that AVs provide both high-quality decision-making
and rapid response times~\cite{gog2021pylot,tan20efficientdet}.

\begin{figure}
  \centering
  \includegraphics[width=0.85\columnwidth]{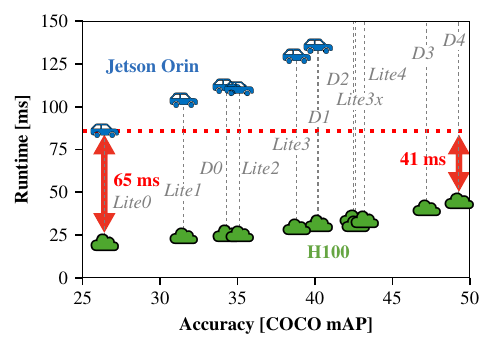}
  \vspace{-1em}
  \caption{
  Cloud accelerators can execute more accurate models with lower
  runtime than hardware designed for autonomous driving. We plot the
  runtimes of different models in the the
  EfficientDet~\cite{tan20efficientdet} family of object detection models
  when executing on the NVIDIA's edge-focused Jetson Orin and cloud-focused
  H100 GPU. Prior work has shown that more accurate object detectors are
  able to improve driving safety by identifying obstacles at greater
  distances~\cite{erdos,schafhalter2023leveraging}.}
  \label{f:latency-accuracy-tradeoff}
  \vspace{-1em}
\end{figure}

Furthering the challenge of meeting stringent performance requirements, the on-car compute AVs have access to is severely constrained today due to physical power, heat, and stability limits~\cite{driving-michigan}, as well as due to economic realities (\S\ref{s:steering-the-convo}); 
taken together, the result is an order-of-magnitude less compute\footnote{A single SOTA cloud GPU (H100) can perform
over $10\times$ more operations per second than AV-targeted chips like
NVIDIA's DRIVE Orin~\cite{h100-spec,drive-orin-spec}.} available on
state-of-the-art (SOTA) AV hardware compared to cloud, which limits which
models can be run in real-time.
Access to better compute would offer the
opportunity to run larger models with higher accuracy faster, directly
translating to improved safety.

We propose turning to the cloud, which offers on-demand access to SOTA hardware and thus provides the opportunity to run larger models for AVs. 
However, data transfers between AVs and the cloud are a major challenge,
as AVs have tight real-time execution constraints and cellular
networks are highly variable and
performance-constrained~\cite{li2021cellularreliability}.
At first glance, using the cloud for autonomous driving seems unrealistic;
AVs process upwards of 8 Gbps of data~\cite{cruise-roscon,siemens-av-data},
an amount that far exceeds the capabilities of today's 5G cellular networks which target uplink speeds of $100$ Mbps despite a theoretical peak
of $10$ Gbps~\cite{5gdeploymentspecs}.
Furthermore, system designers maintain well-founded skepticism of using
cellular networks in critical real-time
systems~\cite{tao2023cellularchallenges} as AVs require
\textit{guaranteed} completion SLOs to achieve safety, which is at odds
with the reliability afforded by cellular networks.

We find a way forward with the following observations:
\begin{enumerate}[noitemsep, topsep=0pt, left=5pt]
    \item AV control systems are a collection of individual services with their own inputs and SLOs, which thus can be individually offloaded based on available bandwidth.
    \item The speedup from remote execution is so large that remote services can tolerate significant data transfer latency, even when running more accurate and expensive models (\cref{f:latency-accuracy-tradeoff}).
    \item Remote execution can be done opportunistically while running a
      local fallback version of the service on-vehicle according to prior
      work~\cite{schafhalter2023leveraging}. This way we can ensure that the cloud will strictly improve the accuracy and safety over running everything locally.
\end{enumerate}
The challenge that remains is deciding which services to offload as available bandwidth varies widely.

In this paper, we make three core contributions.
First, we make the case that using remote resources for autonomous driving
is not only technically and economically feasible, but trends show that it
is \textit{necessary and inevitable} as model compute requirements grow
exponentially, and cars stay on the road for over a decade
(\cref{s:feasibility}).
Second, we introduce a method to define and derive \textit{service-level}
utility curves that capture service accuracy as a function of bandwidth
consumption in order to allocate scarce network bandwidth (\cref{s:method}).
Finally, we discuss the implications of using remote resources for safety-critical, latency-sensitive computation (\cref{s:discussion}).
We believe that our approach generalizes to applications beyond autonomous
driving, and hope our insights will spur further conversations about
the potential of networks to benefit real-time intelligent systems.

\section{How Autonomous Driving Works}
\label{s:background}

\peter{Discuss how each component benefits from cloud.}
\peter{Discuss remote intervention? Might fit better in feasibility.}

At a high level, an AV captures information about its surroundings with a
large array of sensors, and processes that data into control commands (\ie
steering, acceleration, and braking).
Given that an AV operates in dynamic and high-speed environments,
processing must be timely and accurate to ensure safety, with
a response time of ${\sim}400$ ms~\cite{wolfe2020rapid}.

The computation in an AV is typically structured as a pipeline where each
component performs a specific task (\eg object
detection)~\cite{gog2021pylot,cruise-roscon}, many of which are ML-based.
Because more accurate ML models are generally more
compute-intensive~\cite{sevilla2022compute,zhai2022scaling} and thus exhibit
higher latency, constructing an AV pipeline requires careful consideration
of the tradeoffs between accuracy and response time shown in
\cref{f:latency-accuracy-tradeoff}.

We observe that this component-pipeline structure lends itself nicely to
viewing each component \textit{as a service}, as their tasks define a
concrete interface and their implementations select from a diverse set of
potential models and algorithms with a latency-accuracy tradeoff.
We present the key components using an ML in an AV pipeline, and discuss
the types of models on which they rely:

\myparagraph{Sensors.}
The AV pipeline begins with an array of high-fidelity sensors to observe
the surroundings; Waymo's $5$\textsuperscript{th}-generation AVs use $29$
cameras, as well as $5$ lidar and $4$ radar systems to generate a
$360\degree$ view of the
world with a range exceeding $300$ meters~\cite{waymo-av-sensors}.
Similarly, Cruise deploys $16$ cameras, $5$ lidars, and $21$ radars on
their AVs~\cite{cruise-report}.
In total, AV sensors generate over $8$ Gb of data every
second~\cite{cruise-roscon,siemens-av-data} and provide unprocessed
snapshots of an AV's surroundings.

\myparagraph{Perception.}
A collection of ML-based perception modules process
the  sensor data streams and fuse this processed data into an
ego-centric map annotated with nearby obstacles, drivable regions, and
traffic laws (\eg signs, traffic lights).
Perception performs several different tasks such as object detection,
object tracking, and lane detection~\cite{gog2021pylot,apollo-baidu} which
use different models and form subservices.
While there is a large range of perception models for autonomous driving
which use different sensor modalities, most of these models employ convolutional neural networks to extract features from images
or lidar point
clouds~\cite{karpathy-keynote-cvpr-wad-2021,yin2021center,tang-lane-detection-review}.

\myparagraph{Prediction.}
Next, the AV uses ML to anticipate the motion of perceived nearby agents
(\eg pedestrians, vehicles, bicyclists) by processing the outputs from
perception.
Prediction models leverage compute-intensive neural networks such as
Transforms to forecast the future positions of nearby agents based on
patterns in their motion and
behaviors~\cite{rhinehart2018r2p2,shi2022motion,ettinger2024scaling}.
\alex{is this an example of a module with small input size but large
compute?}

\myparagraph{Planning.}
Finally, planning generates safe and comfortable motion plans for the AV
from the perceived and predicted state of the world.
While traditional planners employ search
algorithms~\cite{katrakazas,paden-survey} to generate reliable and
explainable motion plans, we anticipate that future planners will use ML
based on trends in motion planning
research~\cite{hu2023planning,teng2023planning-survey,lu2023imitation,gulino2024waymax}.

We select 2D object detection as an example service to examine the tradeoffs of remote resources in-depth. Object detection is a well-studied perception task with a wide variety of open-source models~\cite{tan20efficientdet,carion2020end,resnet,yolov4} and performs the safety-critical task of identifying and locating nearby obstacles by processing images from the AV's cameras.

We focus on the EfficientDet~\cite{tan20efficientdet} family of object detection models.
Each model in the family, takes a different resolution image as input, ranging from $320\times320$ to $1536\times1536$, and spanning from lightweight, low-accuracy to compute-intensive but high-accuracy, as shown in \cref{f:latency-accuracy-tradeoff} and \cref{t:feasibility-table}.
Our design (\cref{s:method}) increases the accuracy of an object detection service by allocating bandwidth to run more accurate models (\eg EfficientDet D3 and EfficientDet D5) on powerful cloud hardware (\ie H100 GPUs) while meeting a tight latency SLO.
As a result, we can improve accuracy over models capable running on AV hardware (\ie EfficientDet D1 on a Jetson Orin), with the goal of benefiting driving safety.

\section{A Pre-emptive Q\&A}
\label{s:steering-the-convo}

We are interested in designing an offloading system that is deployable in the real world taking advantage of the observations made above. However, utilizing the network for mission-critical applications has been viewed with skepticism for valid reasons. We address some of the potential questions that may arise for the reader.

\myspace{}
\myparagraph{Q:} \textit{If there are better SOTA models today, why would AV manufacturers not use them, releasing AVs that are less safe?}

\myparagraph{A:}
While the exact requirements for deploying AVs are subject to
debate~\cite{cmu2017avdeploy}, it is largely accepted that AVs must first
surpass human driving abilities in terms of safety.
We observe that we are at the cusp of surpassing this threshold as Waymo, Cruise, and others are rolling out extensive public betas across 
several U.S. states~\cite{waymo-los-angeles,cruise-austin,waymo-scaling-to-four-cities}.
These AVs are already safe enough to be on the road; however, we propose
that remote resources provide additional capabilities beyond this minimum
threshold \eg by running SOTA models for which the on-vehicle compute is
insufficient.

\myspace{}
\myparagraph{Q:} \textit{Why not put better hardware on the car if it exists today?}

\myparagraph{A:}
Better hardware is indeed available today, but is
constrained to datacenters for several reasons.
First, cars have stringent power and cooling
constraints~\cite{driving-michigan} whereas SOTA GPUs have notoriously high
power consumption and generate large amounts of heat~\cite{h100-spec,
nvidia-a100-spec}.
Second, AV deployment is inherently an economic problem,
and the cost of new computing hardware is skyrocketing.
In 2024, an NVIDIA H100 GPU costs
\$40k~\cite{2024-h100-price} while a new Tesla costs
\$30k~\cite{tesla-model-3-price}.
Though the pricing and hardware for AV manufacturers is kept secret,
the price of publicly available AV hardware is up to $10\times$
lower than SOTA cloud hardware (\cref{s:feasibility-cost}).

Furthermore, cars are well-suited to statistical multiplexing:
a typical U.S. car owner averages 60.2 minutes of driving per
day~\cite{trafficsafety2023drivingmin}, leaving the car unused for 
${\sim}23$ hours.
Despite correlations in vehicle traffic, there are significant opportunities
to share a pool of compute among many cars, and can be
multiplexed with non-AV workloads when load is low.

However, using the cloud does not exclude improving on-vehicle compute.
If highly resource-efficient hardware becomes available, AV vendors
can and should deploy it on their cars.
What we propose increases flexibility: AV vendors can choose where on the
spectrum they want to operate, and when to invest in improving on-vehicle
or remote resources.

\myspace{}
\myparagraph{Q:} \textit{Isn't the cost of hardware quickly decreasing?}

\myparagraph{A:}
Yes, but model sizes are rapidly
increasing~\cite{sevilla2022compute,liu2022convnet,vit}, as are input
sizes as AVs expand the number and fidelity of their
sensors~\cite{waymo-4th-gen-specs,waymo-av-sensors}.
Furthermore, vehicles have long lifespans:
in 2023, the average age of lightweight vehicles driven in the U.S. is
12.5 years~\cite{bureauoftransportation2024age}.
Even if there is a huge leap forward in compute cost, power, and heat efficiency, millions of older vehicles will remain on the road and will require some way to benefit from new capabilities unlocked by larger models.

\myspace{}
\myparagraph{Q:} \textit{Why not upgrade the hardware on older cars as
costs lower?} 

\myparagraph{A:}
In practice, this depends on the AV deployment model.
Companies that operate fleets of AVs can perform upgrades
as they own, operate, and
maintain all of their vehicles.
However, upgrades for personal vehicles are not as easy.
As a comparison, recalls are mandatory upgrades that fix critical safety risks
such as faulty brake systems~\cite{honda2023recall}.
Though recalls are \textit{free} to the consumer, current fix rates are 52-64\%~\cite{nhtsa2023recalls}
as many car owners do not service their vehicles due to service duration, required
use, and distance from servicing~\cite{cmt-recall-challenge}.

\myspace{}
\myparagraph{Q:} \textit{Why not compress models to be smaller to be able
to run on the hardware that is in deployment?}

\myparagraph{A:}
Model compression is an active area of
focus~\cite{polino2018model-compression,choudhary2020model-compression-overview} 
which typically reduces accuracy to lower compute requirements.
This is orthogonal to our approach as compressed models can still be
too big or too slow to run on an AV.

\myspace{}
\myparagraph{Q:} \textit{How can we rely on anything on the cloud? What if
the network is unreliable, or the cloud fails in some other way?}

\myparagraph{A:}
We emphasize that the network should only be used greedily to improve
performance — existing pipelines running on the car have surpassed human
safety limits already, and the cloud will further improve car performance
where possible.

\myspace{}\noindent
In summary, we present a clear need for remote resources to run the most accurate models for self-driving.
While this need is immediate, there are valid concerns around the technical and economic feasibility of using the cloud for safety-critical, real-time computation, which we address in \cref{s:feasibility}.

\peter{Cut this paragraph?}

\section{Feasibility: A Path Forward}
\label{s:feasibility}

\begin{figure*}
    \centering
    \includegraphics[width=.95\textwidth]{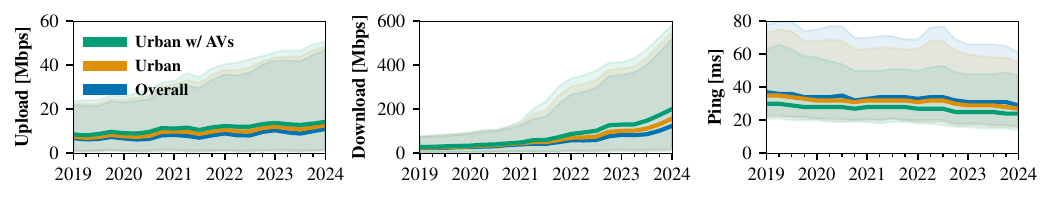}
    \vspace{-1.5em}
    \caption{
      User-experienced
      cellular network speeds are improving according to Ookla's
      SpeedTest.net dataset~\cite{ookla-network-performance}. We find
      improvements in upload, download, and ping speeds across the
      10\textsuperscript{th}, 50\textsuperscript{th}, and
      90\textsuperscript{th} percentiles between 2019 and 2024, capturing
      the expansion of 5G in 2019 and the shut-down of 3G in 2022.
      As the data aggregates connections ranging from 2G to 5G, we
      are unable to report performance by network generation.
      }
    \label{f:speedtest-results}
    \vspace{-0.5em}
\end{figure*}

Latency requirements and AV deployment economics raise questions on the feasibility of using remote resources.
We examine these constraints and address these questions in light of
network and compute trends today.

\subsection{Performance}
\label{s:feasibility-performance}
The stringent performance requirements discussed in \cref{s:background}
lead to the following questions:
(1) do cellular networks support data-transfer latencies low
enough to access remote resources in real time, and
(2) are cloud compute runtimes fast enough to make up for this
network latency?

\myparagraph{Network.}
While cellular rollouts over the past decade have promised great peak
performance, real-world experience of ping and bandwidth is highly
variable which stresses the connection to remote resources.
5G deployment~\cite{marvell2020cellulardeployment} supports 300 Mph speeds, $<$1 ms tower RTT, and 2.6 million devices per square mile~\cite{itu20225g6gcapabilities}.
While the peak maximum total bandwidth provided by 5G connections is
20 Gbps~\cite{qualcomm5greport}, real-world experience varies greatly as
deployments are (1) subject to real-world constraints such as
interference~\cite{siddiqui2021-5g-interference} and (2) configured to
match operator preferences on power, cost, and uplink/download bandwidth
allocation.
We analyze Ookla's Speedtest.net dataset~\cite{ookla-network-performance}
and find that real-world cellular networks provide
over $10\times$ more downlink than uplink bandwidth, and
consistent latencies despite improvements in bandwidth.
5G deployments target a minimum (5th-percentile) user-experienced bandwidth of 50 Mbps for uplink and
100 Mbps for downlink~\cite{5gdeploymentspecs}

\begin{table*}
\centering
  \small
  \begin{tabular}{|c|c|c|c|c|c|c|c|}
\multicolumn{3}{c|}{\textbf{Model}} & \multicolumn{2}{c|}{\textbf{Compute}}
  & \textbf{Network} & \multicolumn{2}{c}{\textbf{Total}} \\
\hline
\textbf{Name} & \textbf{Accuracy} & \textbf{Input Size}
  & \textbf{Orin [ms]} & \textbf{H100 [ms]} & \textbf{Transfer [ms]}
  & \textbf{H100 $+$ Transfer [ms]} & \textbf{Speedup} \\
  \hline ED0 & $34.3$ & $512 \times 512$ & $112$ & $26$ & $20$ & $46$ &
  $2.4\times$ \\
  \hline ED1 & $40.2$ & $640 \times 640$ & $136$ & $32$ & $24$ & $56$ &
  $2.4\times$ \\
  \hline ED3 & $47.2$ & $896 \times 896$ & $325$ & $41$ & $36$ & $77$ &
  $4.2\times$ \\
  \hline ED5 & $51.2$ & $1280 \times 1280$ & $1067$ & $65$ & $52$ & $117$ &
  $9.1\times$ \\
  \hline ED7 & $53.4$ & $1536 \times 1536$ & $1955$ & $101$ & $83$ & $184$
  & $10.6\times$ \\
\hline
\end{tabular}

\caption{Network and compute time for a selection of object detector models
  \cite{tan20efficientdet} with varying input resolutions.
  We compute network time using a bandwidth of 200 Mbps and an RTT of 12 ms
  based on prior measurements~\cite{narayanan20215g-performance}.
  }
\label{t:feasibility-table}
\vspace{-2em}
\end{table*}

We analyze how network conditions impact the runtime of models running in
the cloud from the perspective of an AV in \cref{t:feasibility-table},
assuming a 
5G connection to a nearby datacenter with an RTT of $12$ ms and
an upload bandwidth of $200$ Mbps based on prior
5G measurements~\cite{narayanan20215g-performance}.
We find that improvements in inference runtime outweigh the network
latency for variants of the EfficientDet object detection model (\cref{f:latency-accuracy-tradeoff,t:feasibility-table}), demonstrating that current cellular networks provide
sufficient performance for remote execution.

\myparagraph{Compute. }
We now compare the performance of on-vehicle and cloud compute.
The center columns in \cref{t:feasibility-table} provide the runtime of
several variants of the EfficientDet (ED) object detection model when executing
on NVIDIA's Jetson Orin, an edge-focused chip for ML inference\footnote{The
Jetson Orin uses the same system on a chip as NVIDIA's DRIVE Orin which is
designed to power autonomous driving
capabilities~\cite{nvidia-drive-faq}.},
and NVIDIA's H100, a cloud-focused GPU with top performance on inference
tasks according to the MLPerf Inference: Datacenter
benchmark~\cite{mlperf-inference-leaderboard}.
The H100 executes ED models $4-19\times$ faster than the Jetson Orin, making previously-infeasible models capable of running within SLO, even when factoring in network time (\cref{s:method}).

\subsection{Cost}
\label{s:feasibility-cost}

Given we have shown that network and compute performance is sufficient to
make this approach technically feasible, here we examine whether it is
\textit{economically} feasible.

\myparagraph{Network.}
Commercial network usage is primarily charged by the
GB~\cite{zipit2023wirelesscost}.
Consumer-facing plans vary in price~\cite{cable2023averagecellularcost} from as low as \$0.001/GB in Israel to
over \$2 in Norway. %
This wide range in pricing requires careful consideration in deployment:
\cref{t:network-costs} shows the cheapest \textit{consumer}-facing (\ie SIM
card) cost per GB of data in a selection of countries, along with the
computed cost per hour of streaming 50 Mbps of data.
We note that we expect wholesale pricing, especially geofenced to a
particular region, to be considerably cheaper.

In countries such as Israel, the price of cellular data transmission is
trivial at \$0.02 per hour of driving.
In other countries, including the U.S. price of \$16.88 per
hour, prices are considerably higher and present an economic obstacle to
using remote resources.
In countries with high cellular data prices, operators may choose to reduce
costs by selectively utilizing remote resources to aid in high-stress
driving environments \eg during poor visibility due to weather and busy
urban areas.
\peter{Mention models with small input sizes?}
We further observe a strong downwards trend in the cost of cellular data:
from 2019-2024, the median price per GB worldwide decreased $4\times$ from
\$5.25 to \$1.28~\cite{cable2023averagecellularcost}, and we expect prices to continue to drop.

\begin{table}
\vspace{-1.5em}
\centering
\small
  \begin{tabular}{| c  |c  |c  |c  |c  |}
\hline
\textbf{Rank} & \textbf{Country} & \textbf{\$/GB} & \textbf{\$/Hour}\\
\hline
1 & Singapore & \$0.07 & \$1.65 \\
\hline
2 & Netherlands& \$0.36 & \$8.04 \\
\hline
3 & Norway & \$2.09 & \$47.07 \\
\hline
4 & United States& \$0.75 & \$16.88 \\
\hline
5 & Finland & \$0.26 & \$5.81 \\
\hline
-- & China & \$0.27 & \$6.14 \\
\hline
-- & Israel & \$0.001 & \$0.02 \\
\hline
-- & \textit{10th pct} & \$0.062 & \$1.39 \\
\hline
-- & \textit{Median} & \$0.37 & \$8.42 \\
\hline
\end{tabular}

  \caption{Network costs ranked highest by AV readiness
  score~\cite{kpmg-self-driving-report}. We include China as a major
  AV market~\cite{mckinsey2024avleaders}, Israel as the cheapest cellular
  market, and the 10th percentile and median global country by network
  price. Hourly rates assume a network utilization of 50 Mbps.
  }
\label{t:network-costs}
\vspace{-3.2em}
\end{table}

\noindent\textbf{Compute.}
Cloud providers offer competitive access to GPUs: Lambda Labs hourly pricing ranges from \$0.80 for an NVIDIA A6000 GPU to \$2.49 for an NVIDIA H100 GPU~\cite{lambda-labs-gpu-pricing}.

Sharing compute costs across a fleet of vehicles presents a significant
opportunity to reduce costs even further compared to installing dedicated compute
hardware in each car.
The average driver in the U.S. drives only 60.2 minutes per
day~\cite{trafficsafety2023drivingmin}, \ie a vehicle utilization of
$4.2$\%.
This underutilization is more pronounced for personal vehicles than
autonomous ride-hailing services, for which we expect utilization to be
${\sim}59$\% based on the ratio of peak to average hourly Uber rides in New
York City~\cite{uber-nyc-pickups}.
Considering this underutilization, the cost of purchasing a single H100 GPU
(${\sim}\$40$k~\cite{2024-h100-price}) is equivalent to renting an H100 in
the cloud for an 44 years for the average American driver, and 3 years for
an average autonomous ride-hailing vehicle.

Cloud compute offers a number of advantages.
Cloud access allows operators to configure which compute resources to
use based on compute requirements and cost sensitivity.
Remote resources cannot be stolen or damaged in an accident.
Furthermore, model serving systems can optimize resource utilization by
batching and scheduling requests~\cite{clipper,infaas,clockwork}, resulting
in further price improvements.

\myparagraph{Total cost.}
We find remote resources are cost-effective for personal vehicles and
cost-competitive for ride-hailing.
We estimate total hourly cost of remote resources at
\$3.88, with \$1.39 from the network\footnote{
Mobile networks are cost-effective at the 10th percentile of global
prices which includes major markets such as India, Italy, and Israel.
    We use the 10th percentile of global prices  as an estimate because (1) we expect most of the
world to follow downwards price trends as seen in Israel, and (2)
wholesale, regional pricing for AV providers may be significantly cheaper.
}
and \$2.49 from compute for an H100.
We emphasize that the true cost of cloud compute is likely
lower due to better efficiency when operating at scale.
Considering utilization, we find that a personal AV can operate for 28
years and a ride-hailing AV for 2 years, before matching the
cost of purchasing an H100 GPU (${\sim}\$40$k), indicating that
remote resources are economically viable.

\section{Utility-Informed Method}
\label{s:method}

The goal of our approach is to select which subset of the services that make up the AV control pipeline to run in the cloud, and how to allocate bandwidth between them.
We use perception models, particularly the EfficientDet (ED) family~\cite{tan20efficientdet} described in \cref{s:background}, as our guiding example.
Each service may have multiple models that can run in the cloud.

To achieve this goal, we consider three key dimensions:

\noindent \textbf{(1) Latency SLOs} are set for each service in the AV pipeline. For object detection, we select an SLO of 150 ms based on prior work~\cite{erdos}.
A service's ability to meet its SLO depends on the runtime of the model on the chosen hardware and, for remote services, the data transmission time.

\noindent \textbf{(2) Bandwidth} varies over time (\cref{s:feasibility-performance}), and dictates how quickly data is transmitted. Allocating higher bandwidth to a service allows its input to arrive to the cloud faster, leaving more time for compute within the latency SLO envelope. Given that downlink bandwidth is ${\sim}10\times$ higher than uplink~\cite{5gdeploymentspecs, narayanan20215g-performance} and the output of object detection is relatively small 
(${\sim}4$ KB),
the majority is propagation time (ping RTT) and uplink transmission time (bandwidth $\times$ input size).

\noindent \textbf{(3) Accuracy}
measures how well an ML model performs its task, and is typically computed offline on a pre-defined test dataset (\eg COCO~\cite{coco}).
Object detection uses mean average precision~\cite{pyimagesearch2022cocomap} (mAP) as an accuracy metric, which captures a model's ability to correctly identify objects and place their bounding boxes.

To capture the interaction of these concepts, we extend prior work~\cite{zegura1999utility}
which creates \textit{utility curves} that represent how allocated bandwidth provides utility to an application.
We use service accuracy as a proxy for utility to an AV\footnote{We note that predicting each component's contribution to overall AV driving accuracy is an open problem~\cite{gog2021pylot,philion2020learning}, and is typically approximated by extensive testing of the end-to-end AV pipeline in simulation~\cite{fremont-pldi19}.}
which increases as sufficient bandwidth is made available to run more accurate models.

\begin{figure}
    \centering
    \includegraphics[width=0.8\columnwidth]{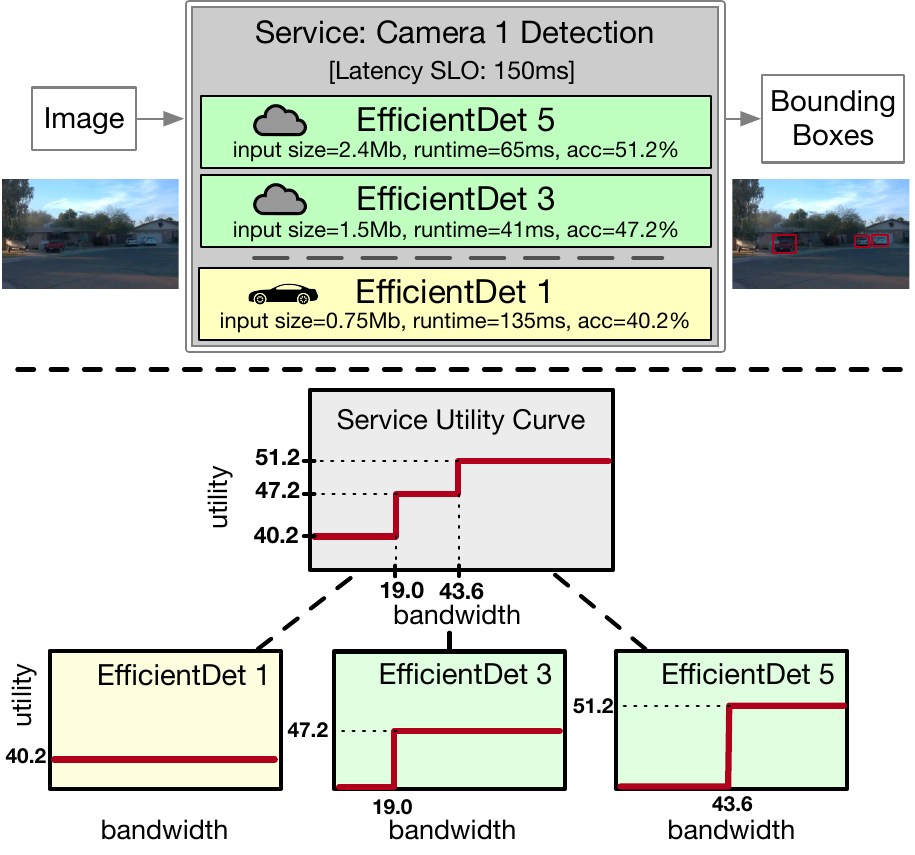}
    \vspace{-0.5em}
    \caption{Utility curves for a service with 3 object detection models.
    }
    \label{f:utility-curves}
    \vspace{-1em}
\end{figure}

We show the construction of a utility curve in \cref{f:utility-curves} for an object detection service using EfficientDet D1 (on-car) and ED3/ED5 (cloud).
As discussed in \cref{s:background}, these are production-class detection models which take different image resolutions and provide distinct latency-accuracy trade-offs. In the ED family, ED1 is the most accurate model that can run on-vehicle within the latency SLO. ED3 and ED5 provide higher accuracy, but are only capable of running within the latency SLO on datacenter hardware; this requires sufficient bandwidth to transfer inputs to the cloud. If available bandwidth is too low, network latency is too high and the SLO is violated, making the utility is 0.

\setlength{\abovedisplayskip}{5pt}
\setlength{\belowdisplayskip}{5pt}
Each cloud model has a tipping point at which the bandwidth $B$ is sufficient to meet the service's latency SLO $t_{\text{SLO}}$ given the RTT $t_{\text{RTT}}$, execution time $t_{\text{exec}}$, and input size $s$:
\begin{equation*}
    t_{\text{SLO}} > t_{\text{total}} = \frac{s}{B} + t_{\text{RTT}} + t_{\text{exec}}
\end{equation*}
Consequently, the utility function for a single model is a step function, with the step from $0$ to the model accuracy at:
\begin{equation*} 
B_{\text{step}}= s / (t_{\text{SLO}} - t_{\text{RTT}} - t_{\text{exec}})
\end{equation*}

For ED3 and ED5, using the numbers in \cref{f:utility-curves}, the tipping point bandwidths are 19.0 Mbps and 43.6 Mbps respectively, resulting in the curves at the bottom of \cref{f:utility-curves}. The utility curve for ED1 is flat as it runs on-car and always meets the latency SLO regardless of allocated bandwidth.

To construct the service-level bandwidth utility curve, we compose all individual model utility curves. As we choose to run only the best feasible cloud model at a time, we take the \textit{maximum} utility across all curves as shown in \cref{f:utility-curves}. This composed curve provides the maximum utility (\ie accuracy) our service can achieve given an amount of bandwidth, while the on-car model provides a guaranteed minimum.

\noindent \textbf{Allocating Bandwidth.}
Given a collection of \textit{services} with utility curves, these curves can be used to formulate a utility maximization problem for allocating bandwidth across all services with any operator-specified policy constraints. 
The utility maximization problem may be converted to an integer linear program~\cite{lasdon2002optimization} for which solvers such as Gurobi~\cite{gurobi} can find optimal bandwidth allocations.
Prior work proposes max-min bandwidth allocation methods using utility curves~\cite{zegura1999utility, google2015bwe}.
We leave further discussion of these methods to future work.

\section{Discussion}
\label{s:discussion}

We conclude with takeaways to guide further research on using remote
resources for autonomous driving.
We address both the implications of our proposed approach as well as
high-level conclusions.

\peter{Another open problem is bandwidth partitioning (slicing total
bandwidth -- space) vs. ordering message transmission (time).
I think the explanation require nuance, so I've dropped it for now.
}

\myspace{}
\myparagraph{Takeaway 1:}
\textit{Dynamic driving environments require dynamic utility
curves.}
AVs navigate a wide variety of driving environments which affect
the performance of a model.
Changes in weather, time of day, and driving location create a
``distribution
shift''~\cite{malinin2021shifts,filos2020av-distribution-shifts}
which may cause model accuracy to degrade if the observed data differs
from the training dataset. 
This will necessitate dynamic utility curves, requiring AVs to monitor performance and update bandwidth allocations to respond to changing driving environments.
We underscore that dynamic utility curves also unlock the opportunity to
use specialized models optimized for specific environments, such as
driving at night, which improves driving safety.

\myspace{}
\myparagraph{Takeaway 2:}
\textit{Bandwidth can be partitioned across space and time.}
\peter{Consider cutting if we are running low on space.}
While we consider services that transmit data simultaneously,
bandwidth may also be partitioned across time.
In this model, services receive time slots in which they may transmit data
using all available bandwidth.
Optimizing the order of service transmission can minimize latency, which
has similarities to prior work in scheduling~\cite{tetrisched,clockwork}.

\myspace{}
\myparagraph{Takeaway 3:}
\textit{Deployments at scale will increase resource contention
and generate new opportunities.}
As more AVs rely on remote resources, they will place a greater load on cellular networks which is magnified by correlations in driving patterns (\eg rush hour).
Network outages from excess load already impact AVs~\cite{cruise-outside-lands-outage} highlighting the need for fleet-wide bandwidth allocation.
Scale also affects our model of remote resources.
In this work, we adopt an ``infinite resources'' view of the cloud.
In practice, a fleet of AVs may exhaust cloud resources which raises the
question of how to model the availability of remote resources: as infinite,
a shared pool, or a static allocation per vehicle.
These considerations affect how AVs use remote resources and may impact the
utility functions.
While scale stresses the available resource, scale also provides an
opportunity to benefit safety via collaboration \eg by sharing data to
reduce blind spots~\cite{zhang2021emp,schafhalter2023leveraging}.

\myspace{}
\myparagraph{Takeaway 4:}
\textit{Cellular bandwidth is the primary source of latency.}
While 5G targets device-to-tower latencies as low as 1
ms~\cite{5gdeploymentspecs}, uplink bandwidths of around 100-200
Mbps~\cite{narayanan20215g-performance} cover only a fraction of the
estimated $8$ Gbps of data AVs produce.
At these rates, transmitting 1 Mb of data takes 5-10 ms.
As backhaul bandwidth is large~\cite{tipmongkolsilp2010cellbackhaul},
device-to-cell-tower bandwidth is the bottleneck.
This leads to the remarkable implication that co-locating compute resources
with cell towers has a relatively minor impact on latency, indicating
deploying resources in nearby datacenters is sufficient as increasing
device-to-server distance from 0 km to 320 km increases latency from 6 ms
to 12 ms~\cite{narayanan20215g-performance}.

\myspace{}
\myparagraph{Takeaway 5:}
\textit{Remote resources are cost-effective.}
We re-visit the economic trends which reveal that remote
resources are cost-effective (\cref{s:feasibility-cost}).
While the price of data transfers varies wildly from \$0.001-\$2.00 per GB (\cref{t:feasibility-table}), network prices are rapidly decreasing with the the median price worldwide dropping $4\times$ between 2019 and 2024.
In contrast, compute prices are high at around \$2.49 per hour for a remote
H100 GPU compared to an amortized cost of \$2.28 per hour when deploying an
H100 on-vehicle.
However, the low utilization of vehicles makes remote resources more
cost-effective, as idle remote resources can be re-purposed.

\myspace{}
Remote resources have the potential to improve driving safety by running
more accurate models more quickly.
While reliability of the network remains a concern that must be mitigated
by on-vehicle fault tolerance methods, existing 5G deployments provide
sufficient bandwidth and latency to access remote resources.
We find that bandwidth is the limiting factor to using remote resources,
and present a design that uses utility curves to allocate bandwidth across
the services comprising an AV.
We argue that these trends demonstrate that the use of remote resources is
necessary and inevitable to maximize driving safety, and believe that
addressing this challenge opens exciting new opportunities for research.

{
  \bibliographystyle{plain}
  \IfFileExists{bibliography/articles.bib}{
      \bibliography{bibliography/articles,bibliography/books,bibliography/mine,bibliography/papers,bibliography/patents,bibliography/standards,bibliography/techreports,bibliography/theses,bibliography/urls}
  }{
      \bibliography{../bibliography/articles,../bibliography/books,../bibliography/mine,../bibliography/papers,../bibliography/patents,../bibliography/standards,../bibliography/techreports,../bibliography/theses,../bibliography/urls}
  }
}

\end{document}